# Direct transfer of light's orbital angular momentum onto non-resonantly excited polariton superfluid

Min-Sik Kwon[1,2,†], Byoung Yong Oh[1,†], Su-Hyun Gong[1,2,4], Je-Hyung Kim[1,2,∥], Hang Kyu Kang[3], Sooseok Kang[3], Jin Dong Song[3], Hyoungsoon Choi[1*], and Yong-Hoon Cho[1,2*]

[1] Department of Physics, Korea Advanced Institute of Science and Technology (KAIST), Daejeon, Republic of Korea

[2] KI for the NanoCentury, Korea Advanced Institute of Science and Technology (KAIST), Daejeon, Republic of Korea

[3] Center for Opto-Electronic Convergence Systems, Korea Institute of Science and Technology (KIST), Seoul, Republic of Korea

[4] Department of Physics, Korea University, 45 Anam-ro, Seongbuk-gu, Seoul, 02841, Republic of Korea

† : These authors contributed equally to this work.

*: These authors are corresponding author who contributed equally.

e-mail : h.choi@kaist.ac.kr; yhc@kaist.ac.kr

[∥] Present address: Department of Physics & School of Natural Science, UNIST, Ulsan 44919, Republic of Korea

**Recently, exciton-polaritons in a semiconductor microcavity were found to condense into a coherent ground state much like a Bose-Einstein condensate and a superfluid. They have become a unique testbed for generating and manipulating quantum vortices in a driven-dissipative superfluid. Here, we generate exciton-polariton condensate with non-resonant Laguerre-Gaussian (LG) optical beam and verify the direct transfer of light's**

**orbital angular momentum to exciton-polariton quantum fluid. Quantized vortices are found in spite of large energy relaxation involved in non-resonant pumping. We identified phase singularity, density distribution and energy eigenstates for the vortex states. Our observations confirm that non-resonant optical LG beam can be used to manipulate chirality, topological charge, and stability of non-equilibrium quantum fluid. These vortices are quite robust, only sensitive to the OAM of light and not other parameters such as energy, intensity, size or shape of the pump beam. Therefore, optical information can be transferred between photon and exciton-polariton with ease and the technique is potentially useful to form the controllable network of multiple topological charges even in the presence of spectral randomness in solid state system.**

A quantum vortex, initially discovered in superconductors [1-3], superfluid helium [4-6] and cold atoms [7,8] is a topological defect (resulting from) having rotational superflow with quantized phase winding. Due to the topological stability, it has potential applications in data storage and transfer [9-11]. Fundamentally, a quantum vortex is a signature of phase coherence in a superconductor or a superfluid carrying either a quantized magnetic flux or a quantized angular momentum, respectively. Studying the quantized vortex in exciton-polaritons has advantages over other superfluid and superconductor systems, in the sense that the system includes both photonic component and excitonic component which can be controlled optically and easily visualized [12-16]. A microcavity exciton-polariton has a finite lifetime and decays by leaking photons out of the cavity making the system inherently non-equilibrium [17-20]. The leaking photons carry the density, momentum, energy, spin and phase information of the polaritons. One can characterize the non-equilibrium dynamics of the polariton fluid by characterizing the photo-luminescence [21-23].

Creating vortices in an exciton-polariton superfluid, however, has been technically challenging and a focus of recent researches. Two most effective methods of controlled vortex creation have been utilizing triggered optical parametric oscillation (TOPO) for resonant pumping [21] or creating geometrically engineered beams for non-resonant pumping [24]. TOPO requires a pump beam to be injected at a very specific point on the dispersion curve to induce optical parametric oscillation (OPO) [25,26]. An additional probe beam with an orbital angular momentum (OAM) triggers the OPO process to take place with the resulting condensate having the same angular momentum as the probe beam. The downside of this technique is that it requires two beams one of which must be precisely tuned to have exact energy and momentum.

Geometrically engineered non-resonant pump beams, on the other hand, are based on non-resonant pumping and are free from these constraints. They suffer, however, from a different type of fine tuning. These pumps create a potential landscape from the pump beam. The amount of circular flow cannot be easily controlled as the potential landscape is sensitive to the size, shape, and intensity of the beam.

In this letter, we demonstrate that, for the first time, quantized vortices can be generated by simply transferring OAM of a single pump beam on to polariton superfluid. The most remarkable feature of our result is that the pump beam does not have to be on resonance with the polaritons for optical control, exploring excited states of coherent polariton fluid combined with energy relaxation in out-of-equilibrium system. Current theoretical model, such as generalized Gross-Pitaevskii equation mainly describe dynamics of a coherent polariton condensate in lower polariton branch (LPB) based on mean field theory [27]. This model is coupled with rate equation of exciton reservoir density, which has considered pumping and loss of polaritons with stimulated scattering in LPB. In non-resonant pumping, hot electron-hole plasma with higher energy experiences energy relaxation by phonon emission down to exciton reservoir states. However, the specific and exact description of this energy relaxation has needed more studies including transferring OAM under non-resonant pumping in the range of energy above LPB. The optical OAM information was widely believed to be lost along the energy relaxation process. In the experiment of this letter, however, our results show that OAM of the non-resonant pump beam can be transferred to polaritons. As we will discuss below, this method of vortex generation is significantly robust against a change in energy, intensity, shape, and size of the pump beam. Most of the constraints from the previous methods described above are relieved and a very simple way of controlling quantum vortices is now possible.

In our experiment, the sample consisted of GaAs quantum wells (QWs) and distributed Bragg reflector (DBR) structure forming a microcavity. Ti:Sapphire pulse laser with an energy of 1.73 eV (716 nm, center wavelength) was shone non-resonantly onto the semiconductor microcavity sample in a cryostat at 6K (Fig. 1a). Laguerre-Gaussian (LG) beam was generated from a diffractive optic component. Interference of polaritons was measured with a modified Mach-Zehnder (MZ) interferometer [14,28,29] integrated with a Fourier optics imaging setup [30,31].

Fig. 1b shows the spatial intensity distribution (Fig. 1b left) and interference (Fig. 1b right) of the Laguerre-Gaussian (LG) pump beam with total orbital angular momentum number, $\ell = +1$ and about 30 μm diameter size reflected from the sample. Density and interference of fluorescence are shown in Fig. 1c and 1d for below and above threshold pump power ($P_{th} \sim 2.5$ mW) of polariton condensation, respectively. Above the threshold power, three noticeable effects occurred. First, the spatial correlation region with coherence expanded in the area of polaritons to about 570 μm$^2$ (Fig.1d). In the low polariton density regime (below threshold), a small region of the interference pattern with no anomaly exhibited short range correlation, stemming from the correlation length of the polaritons' thermal de Broglie wavelength. In the higher density regime, the thermal de Broglie wavelength of the polaritons becomes comparable to their average separation [17]. Once a polariton condensate and superfluidity are formed, the range of coherence expands marked by the interference pattern over a larger area in polariton density distribution (Fig.1d).

Second, there is a fork shape dislocation in the interference pattern (Fig.1d). The polariton condensate can build a quantized circulation which carries a phase winding. The singularity at the center of a $2\pi$ phase winding appears as a branch cut in the interference pattern that we observe in Fig. 1d. In other words, a single quantum vortex is generated when the system is excited by a pump with OAM of $\ell = +1$.

Energy-momentum dispersion relation also goes through an abrupt change as can be seen in Fig. 1e,f. Quadratic dispersion of polaritons suddenly collapses into the ground state (1.59 eV) and the first excited state associated with the vortex above the threshold. The change in dispersion marks a clear transition into the condensate. Spatially resolved photoluminescence (PL) reveals a significant feature of this dispersion (Fig. 1g,h). Below the threshold, polaritons are mostly concentrated around the region excited by the pump beam at all energies. Once the polaritons form a condensate, the ground state and the excited state shown in the figure are discretely separated in energy. The excited state polaritons are collected around the pump beam spot and forms vortices. The concentrated polaritons in this region also serve as a potential landscape and trapped condensates are formed inside the ring as the ground state. In effect, one gets a multistate superfluid unique to a driven-dissipative system. The density distributions of the multiple states are overlapped in real space density image. Thus, a consequence of the multistate condensation is that the vortex core appears to be filled due to the ground state near the center of the ring.

To test if the vortex generated in the polariton superfluid is indeed the result of angular momentum transfer from the incident non-resonant laser pump, the comparison between phase winding direction of a laser beam and the resulting vortex was made by utilizing a modified MZ interferometer. In Fig. 2, the MZ interference patterns are shown for both polariton emission (Fig. 2b and 2f) and corresponding pump beam (Fig. 2d and 2h). One can clearly see the phase winding direction of a vortex matching that of the incident laser. This could support that the controlled chirality of the vortex is not the result of the spontaneous breaking of rotational symmetry but that of deterministically broken symmetry depending on the effect of transferred optical OAM. This provides a strong evidence that the angular momentum of an incident laser is transferred to the polariton superfluid. In other words, vortex chirality can be controlled with the OAM of laser even in non-resonant pumping.

We repeated the measurement with different angular momentum values with the pump power fixed at 1.6 times the threshold power (Fig.3). FIG.3a,d,g indicate spatial polariton density distibutions. In order to generate the ring-shaped beam with zero OAM, the incident Gaussian pump beam propagated through Chromium disk blocking mask on glass sheet [32]. When a ring-shaped beam with zero OAM ($\ell = 0$, zero phase winding) was injected, no phase singularity was observed. Diffusive flow pattern can be obtained from the spatial phase distribution (FIG. 3c) The phase map is extracted from the spatial MZ interference (FIG. 3b,e,h). Phase gradient in the radial direction is clearly present as a result of a difference in polariton density along the radius, which creates a diffusive flow. Therefore the polariton density was accumulated around the center of the ring without rotational flow (phase winding), and phase gradient was present both inside and outside the ring as expected.

With an OAM of $\ell = +1$ incident beam, in addition to the radial phase gradient (radial polariton flow), a $2\pi$ azimuthal phase winding is present around the center of the ring (FIG. 3f). When the total OAM of the incident laser was doubled to $\ell = +2$, two phase singularities were seen in the emission (FIG. 3i). Two single quantized vortices were generated, consistent with what was expected for a superfluid. This indicates that not just the direction but also the magnitude of the injected beam's OAM is preserved.

These non-resonantly transferred angular momenta are quite robust in maintaining OAM of polariton vortices even as the pumping intensity is changed. At high intensity, polariton-polariton and polariton-reservoir interactions increase inside the region of pumping beam diameter [33]. Fig. 4 is the phase map extracted from the MZ interferences. Fig. 4a-d shows the phase map of polariton condensates ($\ell = 0$) generated from the ring-shaped ($\ell = 0$) excitation beam. There is no vortex up to about 10 times the threshold pump power. From this, we can comprehend that the interplay between sample inhomogeneity and polariton flow in optically induced potential are negligible in the pumped region of the sample.

For the LG pump beam with the winding number $\ell = +1$, shown in Fig. 4e-h, a single vortex appeared above the threshold (Fig. 4e) and was stably maintained in the center of the LG pump area up to a pump power of ~ 10 $P_{th}$ (Fig. 4f,g). As shown in Fig. 4h, much above the 10 $P_{th}$ pump density, the phase around the single vortex is blurred indicating that the phase fluctuations are present around the vortex, which could eventually become unstable (Fig. 4h. For details, see ref.[34]).

Similarly, when the total OAM of the laser was increased to $\ell = +2$, two vortices were stable up to a pump power of 2 $P_{th}$ (Fig. 4i and 4j). Around 2 $P_{th}$, an additional vortex appeared in the pumped region (Fig.4k). The additional formation of the vortex at high pump power is likely a result of hydrodynamic polariton flow [24] or instability of non-equilibrium polariton condensate [35] based on polariton-polariton and polariton-reservoir interaction in an optically induced potential landscape [36]. Nonetheless, vortices were stable up to a few times the threshold pump power.

As pump power increases very high up to about 30 $P_{th}$, many polaritons can flow radially into the center due to repulsion from the largely populated exciton reservoir located along pumping region through energy relaxation [36] [FIG. S6]. Overlapping of polariton wavefunctions can build coherence in the center. The vortices generated by LG beam pumping can interact with the trapped polaritons at high pump power.

The robustness of the vortices was also checked against the change in pump beam radius. The vortices positioned and maintained their OAM inside pumping area up to 30% increase of the beam radius (for details, see Ref [34]). Even the presence of asymmetry in the intensity of the ring-shaped vortex beam did not affect the vortex formation (for details, see Ref [34]). All of these show that our method of vortex formation by transferring OAM of the non-resonant pump beam is quite robust.

How the orbital angular momentum is transferred in this non-resonant pumping technique remains unknown. Coincidentally, the excited free carriers at 1.73 eV are 140 meV above the polariton ground state of 1.59 eV which is almost exactly four times the longitudinal optical (LO) phonon energy of 36 meV in GaAs. To rule out the possibility of some unidentified angular momentum preserving parametric scattering mediated with LO phonon, we swept our pump laser energy from 1.70 eV to 1.77 eV. Vortices were found when the energy of the pump beam is not an exact integer multiple of the LO phonon energy, suggesting that parametric scattering by LO phonons is unlikely to be responsible for the vortex formation.

What could be happening is a somewhat subtler process. In a driven-dissipative (pumped decay) system, $\ell = 0$ ring-shaped pumping causes steady radial flow away from the ring both inwards and outwards. However, in the regime that the size of pumping spot is larger than that of polariton condensate with Thomas-Fermi radius, the circular symmetry of the flow is predicted to be unstable and non-zero net circulation can be generated [35,37]. The rotational invariance originally present in the pump beam is spontaneously broken in the resulting polariton fluid with the direction of the circulation chosen at random. Our result from the incident beam with $\ell = 0$ shows no such effect, so we are clearly not in this regime. However, injection of orbital angular momentum from LG beam (ring shape) could be inducing this instability that would not be present otherwise. The induced instability (a small residual asymmetry) in exciton reservoir can be enhanced by stimulated scattering in the bottom of LPB. Once the ring shape (flow) symmetric polariton flow becomes unstable, the injected OAM could further nudge the polariton fluid to follow the chirality and OAM of the incident beam.

Alternatively, the multistate nature of the condensate could hold the key to this OAM transfer. The fact that some of the condensates is in the ground state with zero linear and angular momentum indicates that part of the total OAM initially transferred from the laser is lost. If the OAM is completely lost during the relaxation process, the whole system would condense substantially in the ground state. However, as long as some fraction of OAM survives the relaxation process, the quantized nature of the vortices could be forcing the condensate to split into the ground state with zero OAM and the excited states with quantized OAM.

Still, the exact mechanism through which phase information of non-resonant laser is transferred in the polariton superfluid is not well-understood yet and merits a further investigation, for example, parametric scattering mediated with hot exciton or bosonic stimulated scattering as approachable other possible mechanisms. For future studies, advanced theoretical models are needed to describe energy relaxation related to carriers, excitons, and polaritons.

In conclusion, we transferred optical phase winding onto polaritons by using the non-resonant LG pump beam without complicated resonant pump (for example, TOPO) or geometrically engineered non-resonant pump (for example, chiral lenses). Excellent controllability of polariton OAM through a simple manipulation of incident light's OAM is demonstrated, which paves the way for developing non-equilibrium Abrikosov lattice formation and studying its dynamics. The ease of control also provides a new means to study all optical memory devices [38,39], such as vortex memory [40], and simple quantum simulators [41-44].

**ACKNOWLEDGEMENT**

The authors wish to thank L.S. Dang, M. Richard (CNRS, Grenoble, France), I. Savenko (IBS, Daejeon, Republic of Korea), C. Park and M. Kim (KAIST, Daejeon, Republic of Korea) for helpful discussions. This research was supported by National Research Foundation (NRF) of Korea through projects NRF-2016R1A2A1A05005320, 2015R1C1A1A01055813 and 2016R1A5A1008184, and the Climate Change Research Hub of KAIST (Grant No.N11160013). The authors in KIST acknowledge the support from KIST institutional program of flagship.

**Author contributions**

B.Y.O and M.S.K contributed equally to this work. B.Y.O, M.S.K performed the experiment

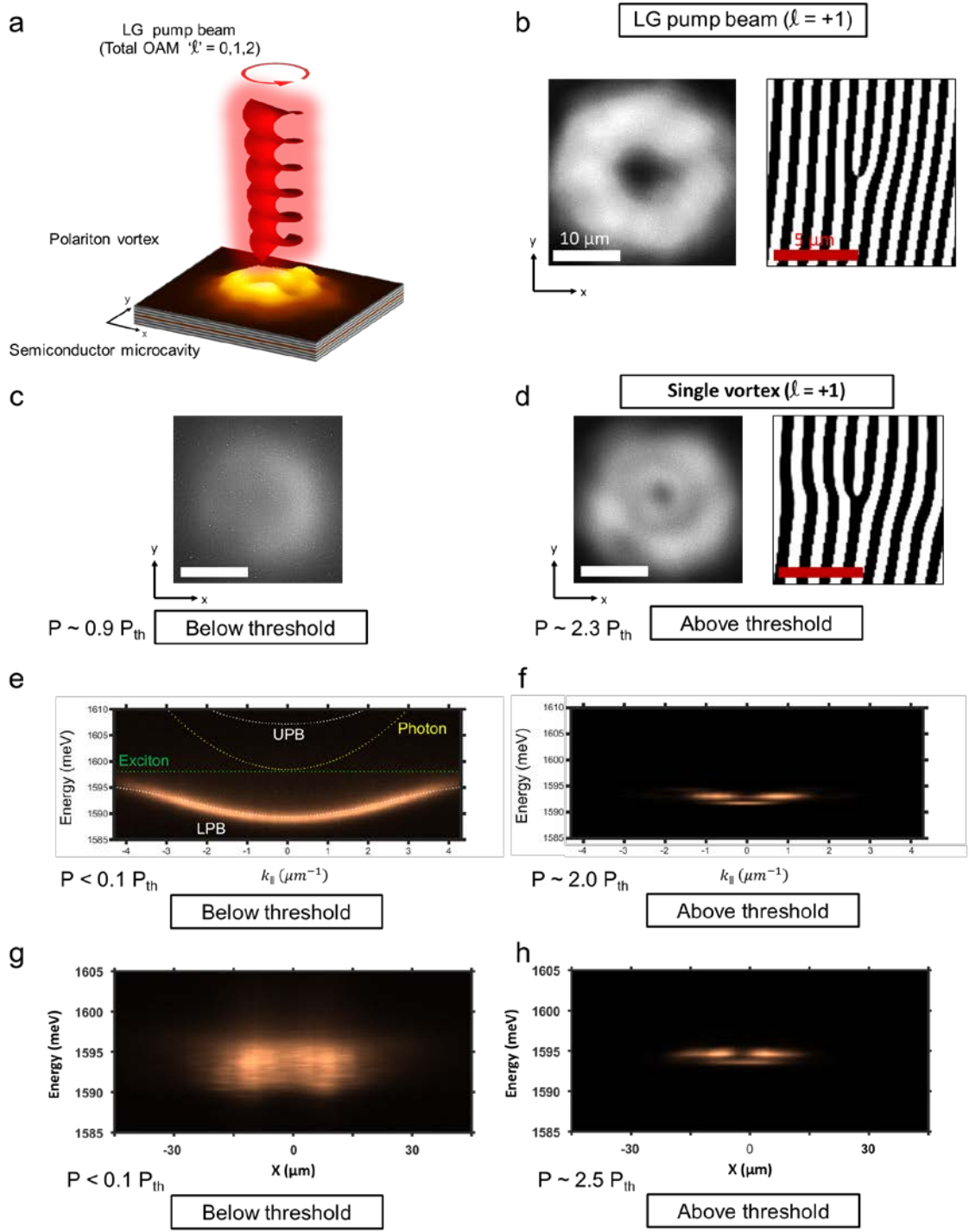

FIG. 1. Creation of a vortex by non-resonant transfer of OAM (phase winding). (a) Schematic configuration of vortex optically pumped with OAM in real space. (b) Spatial distribution of the intensity (left) and spatial interference (right, Mach-Zehnder interferogram) of the LG laser beam with $2\pi$ phase winding ($\ell = +1$). (c,d) Creation of a single quantized vortex with non-resonant LG beam of $\ell = +1$. Spatial density of polaritons below the threshold in c and a single

polariton vortex (white circles) above the threshold pump power in d. Interference "fork" image indicating the presence of the single vortex in d is magnified image extracted by Fourier filtering for clarity. (e) Measured energy-momentum dispersion of lower polariton branch (LPB) below the condensation threshold. White dotted curve, yellow dotted curve and flat green dotted line represent lower and upper polariton branch, cavity photon and exciton dispersion, respectively, obtained through fitting the LPB. (f) The ground and the first excited states of a single vortex, in polariton fluid above the condensation threshold. (g,h) Measured spatial-resolved PL below and above the condensation threshold. (h) Trapped polaritons (ground state) and vortex (excited state) above the condensation threshold in polariton fluid. Scale bars, 10 μm (white), 5 μm (Red)

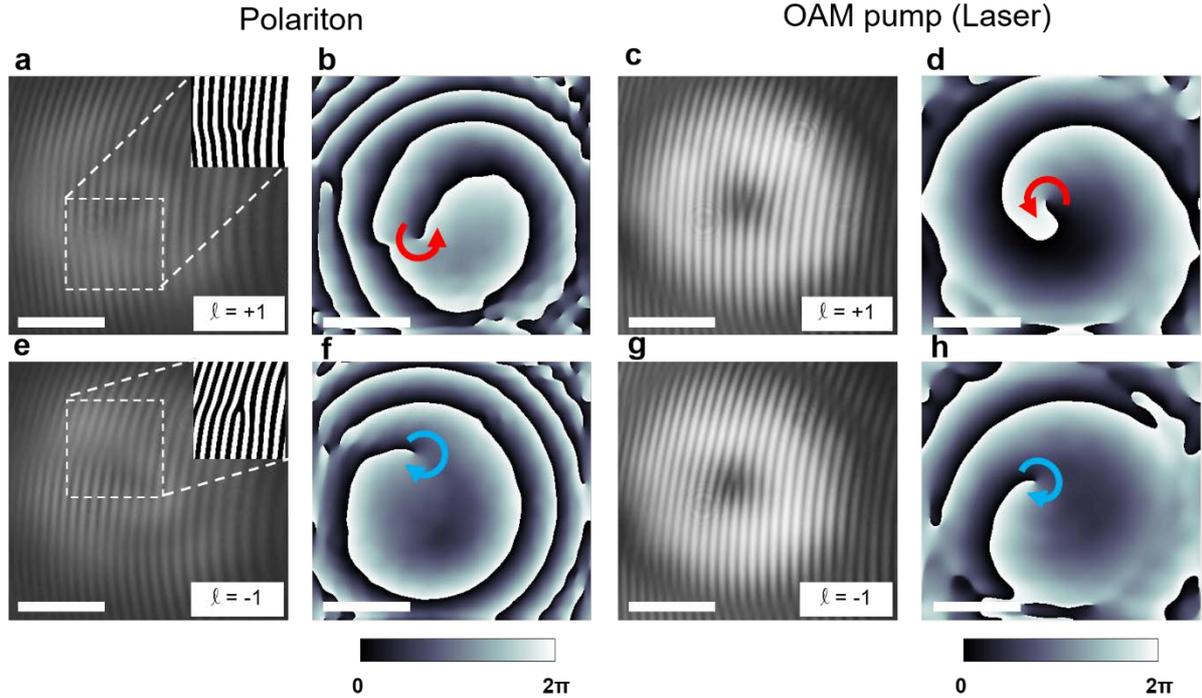

FIG. 2. Polariton vortex chirality vs LG laser beam's chirality (a-d) The incident pump beam has the OAM, $\ell = +1$. (a) Interference image and (b) extracted phase image of the polariton vortex shows a counter clockwise phase winding ($\ell = +1$). (c) Interference image and (d) phase image of the pump beam with the counter clockwise phase winding. (e-h) The same set of images as (a-d) with incident pump beam having $\ell = -1$ OAM. Insets of (a) and (e) were extracted by Fourier filtering the dotted box regions. Scale bars are 10 μm. Curved arrows indicate phase winding directions (chirality) of a single vortex.

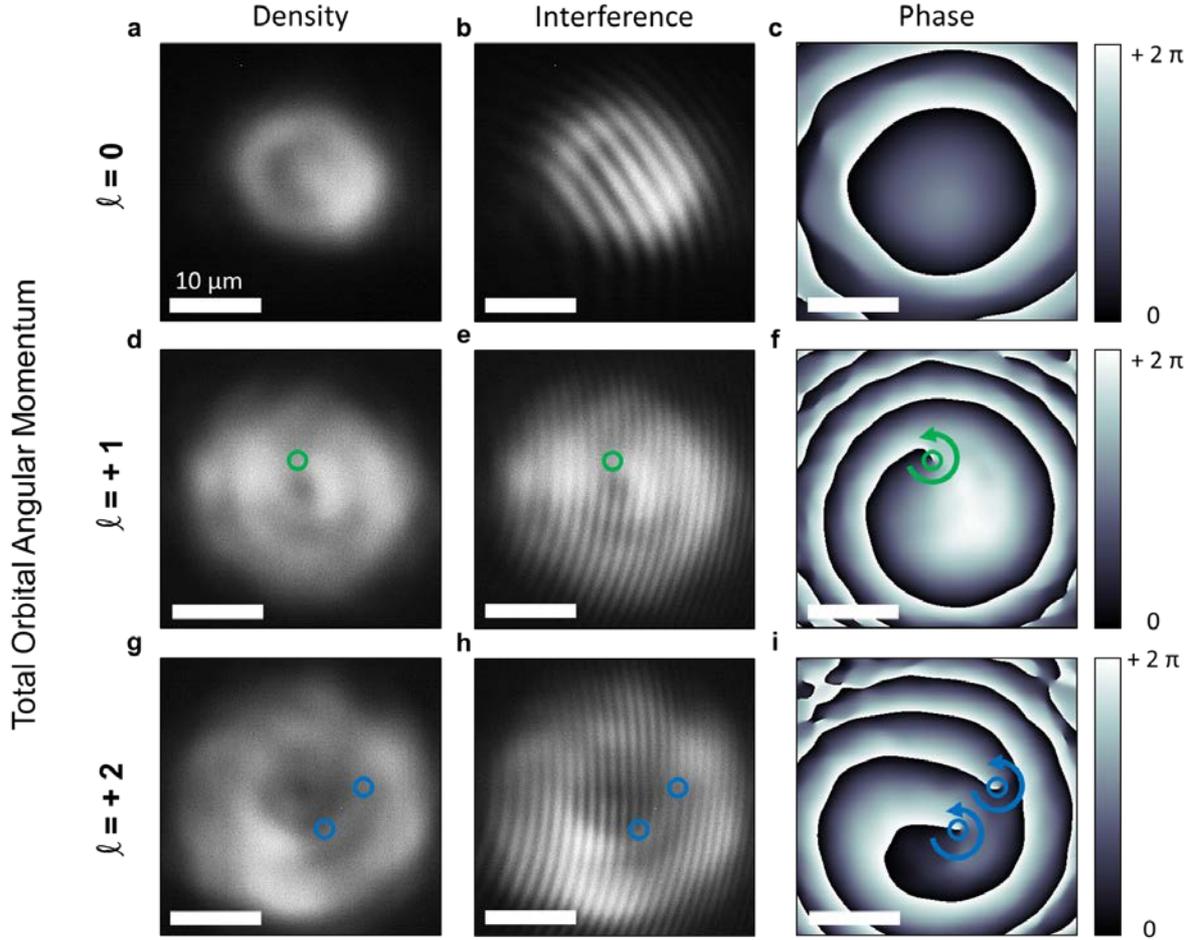

FIG. 3. Pump beam's OAM dependence of the polariton vortex count. OAM of the pump beam is tuned from $\ell = 0$ to $\ell = +2$. (a,d,g) Polariton spatial density distributions. (b,e,h) Spatial MZ interference of polariton condensates. (a,b,c) A ring-shaped beam without OAM ($\ell = 0$) was used. (a) Image of a polariton for $\ell = 0$. In (c) Phase gradient is developed radially with no winding, showing no vortex. (d,e,f) The incident ring beam has the OAM of $\ell = +1$. (d) Image of a polariton emission at $\ell = +1$. (f) a corresponding $2\pi$ phase winding is visible. (g,h,i) The same set of images as (d,e,f) with the difference being the OAM having $\ell = +2$. Two vortices are observed. All of these experiments were carried out at 1.6 $P_{th}$. The white scale bars is 10 μm. Green and blue circles indicate position of vortices. Arrows indicate chirality of vortices.

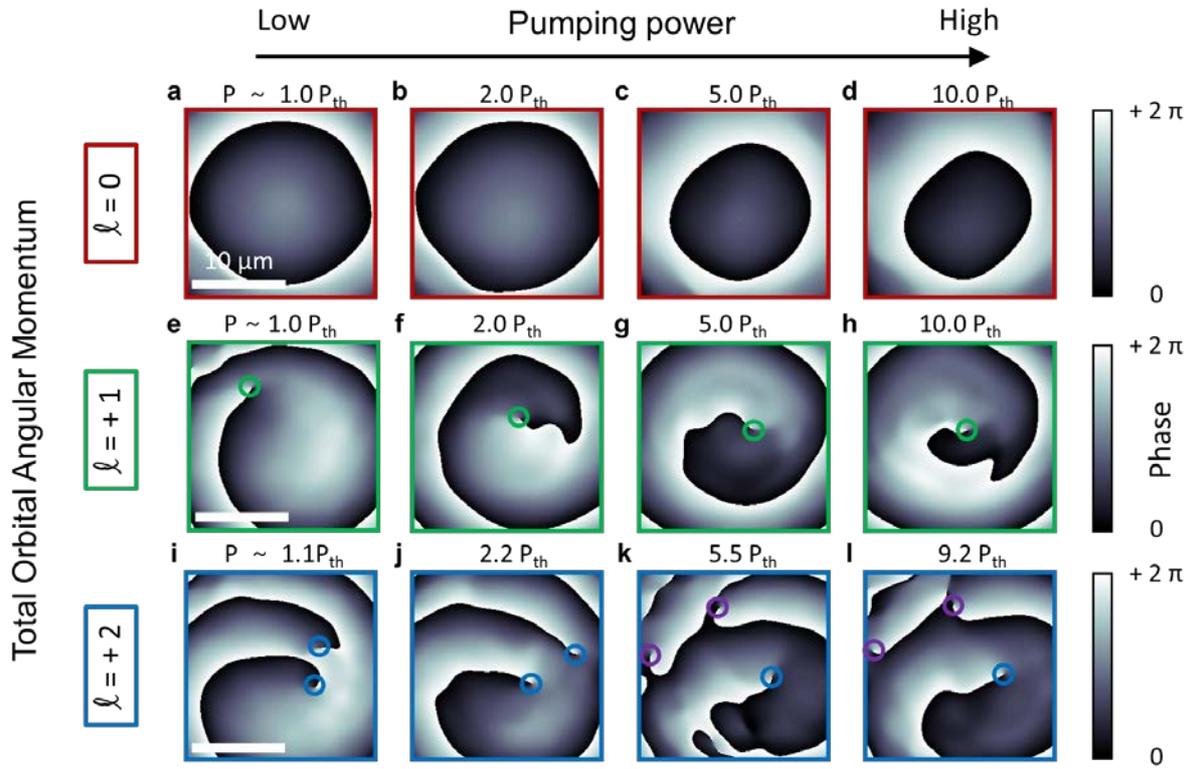

FIG. 4. Stability of polariton vortices against the pumping power. Phase map images of the polariton condensates generated by non-resonant pumping with (a-d) $\ell = 0$, (e-h) $\ell = +1$ and (i-l) $\ell = +2$ are shown for various pumping powers. (a-d) Polariton condensate ($\ell = 0$) with increasing pump power up to 10.0 $P_{th}$ shows no clear vortex. (e-h) A single vortex ($\ell = +1$) is stable up to pump power of 10.0 $P_{th}$. (i-l) Two $\ell = +1$ vortices are stable up to 2.2 $P_{th}$ with $\ell = +2$ LG pump beam. Scale bars, 10 μm.

# Supplemental material

## Direct transfer of light's orbital angular momentum onto non-resonantly excited polariton superfluid

Min-Sik Kwon[1,2,†], Byoung Yong Oh[1,†], Su-Hyun Gong[1,2,4], Je-Hyung Kim[1,2,∥], Hang Kyu Kang[3], Sooseok Kang[3], Jin Dong Song[3], Hyoungsoon Choi[1*], and Yong-Hoon Cho[1,2*]

[1] Department of Physics, Korea Advanced Institute of Science and Technology (KAIST), Daejeon, Republic of Korea

[2] KI for the NanoCentury, Korea Advanced Institute of Science and Technology (KAIST), Daejeon, Republic of Korea

[3] Center for Opto-Electronic Convergence Systems, Korea Institute of Science and Technology (KIST), Seoul, Republic of Korea

[4] Department of Physics, Korea University, 45 Anam-ro, Seongbuk-gu, Seoul, 02841, Republic of Korea

† : These authors contributed equally to this work.

*: These authors are corresponding author who contributed equally.

e-mail : h.choi@kaist.ac.kr; yhc@kaist.ac.kr

∥ Present address: Department of Physics & School of Natural Science, UNIST, Ulsan 44919, Republic of Korea

# S I. SAMPLE AND EXPERIMENT SETUP

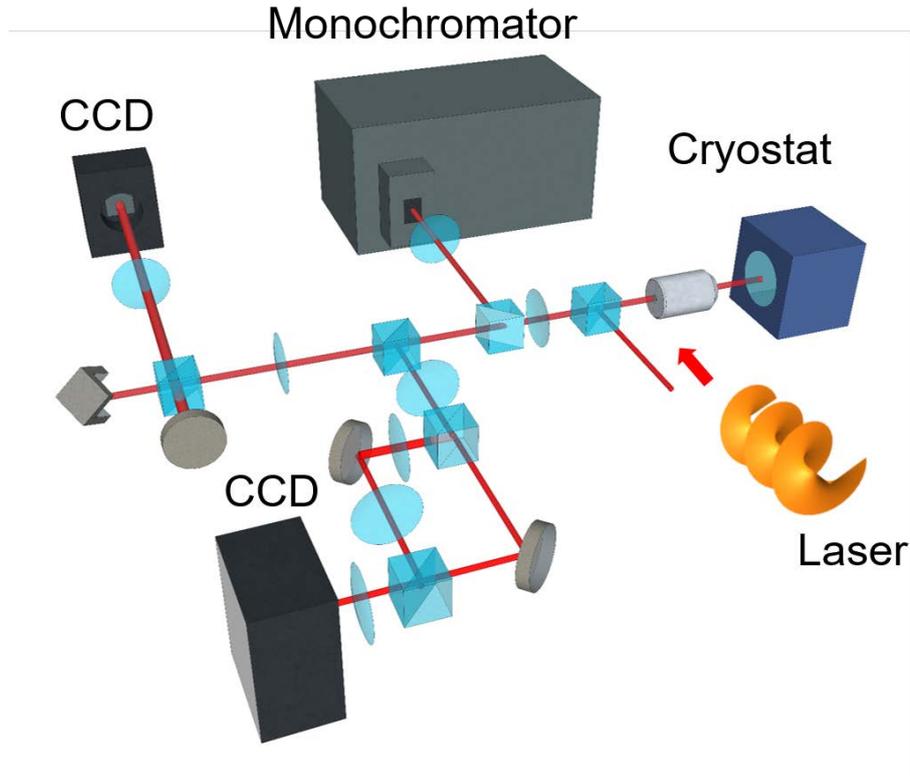

FIG. S1. Schematic of experimental setup. Angle-resolved photoluminescence (PL) with Michelson interferometer and modified Mach-Zehnder (MZ) interferometer. The incident pump beams were pulsed Ti:Sapphire laser beams. Laguerre-Gaussian (LG) beam with orbital angular momentum (OAM, winding number $\ell = +1$ and $+2$) and doughnut beam without OAM ($\ell = 0$) was used. The phase mask converts the Gaussian beam to the LG beam.

A semiconductor sample consisting of QWs, a microcavity, and DBR was made by molecular beam epitaxy (MBE) growth. A $\lambda/2$ AlAs cavity was sandwiched between two AlAs/AlGaAs distributed Bragg reflectors, with 16 and 20 pairs for the top and bottom components, respectively. We have in total 12 QWs in the form of three sets of four QWs. The GaAs 4 QWs were located at the antinodes of the electric field of $\lambda/2$ cavity to increase spatial overlap between the exciton and the cavity photon fields. In order to obtain more coupling of GaAs QW excitons with the cavity field, we added two groups of four quantum wells which are placed at the location of secondary maximum antinodes outsides of the cavity.

The strong coupling regime was confirmed by both reflectivity and E-k dispersion curve measurement. The measurements showed a Rabi splitting of $2\hbar\Omega_R = 17.8$ meV. They indicated cavity photon-exciton detuning $\delta = +0.5$ meV, which are estimated by fitting anti-crossing in the energy momentum dispersion curve.

We performed the experiment with a non-resonant pump scheme, using a mode-locked Ti: Sapphire pulse laser with an energy of 1.73 eV (716 nm) with a 150 fs pulse width and 80 MHz repetition rate. The incident direction of the pump was normal to the sample surface ($k_\parallel = 0$).

For the stable exciton-photon strong coupling regime, it was necessary to decrease the temperature to 6 K using a cryostat. The cryostat had a closed cycle of helium flow and vibration free system. The sample was mounted on a copper sample holder by using silver paste. A piezo stage in the cryostat was used to provide accurate control of position for the sample.

In optical setup, a high numerical aperture (N.A. = 0.55) objective lens was used to make excitation pumping spot on the sample, and collect the emission from exciton-polariton condensate. We obtained the momentum, energy and real space information of polaritons simultaneously using a Fourier microscope combined with a monochromator and charge-coupled device (CCD) camera [1]. To form the OAM of the pump beam, we used a phase mask, whose structure is made of helical phase steps. The phasemask transferred the Gaussian pump beam profile to the ring shaped beam with OAM. For comparing the presence with the absence of transferred OAM in polariton condensates, we made a ring shaped beam without angular momentum ($\ell = 0$) by using a 10 μm chromium mask on a glass plate.

To distinguish between laser and sample luminescence, we used a spectral filter and a dichroic mirror. Because of the filter setup in the measurement section, the laser and luminescence could be distinguished, so that laser intensity was cut to 0.002% of its original intensity.

To clearly observe the location, distribution and phase information of polariton condensates and vortices, an interferometer was installed in the detection setup. Michelson (MC) interferometer was useful in measuring and showing directly spatial correlation ($g^{(1)}$ (r,-r) correlation function) of polariton condensate [2]. In addition, the imaging in MC interferometer of optics setup had more clarity than that of the MZ interferometer which beam intensity decrease in the reference arm by magnification or filtering technique for reference phase beam.

Meanwhile, the Michelson interferometry had one shortcoming in identifying the chirality of the quantum vortices. Two arms of each split beam contained an image of a vortex, one upright and the other retro-reflected. When these beams interfered with each other, the interference pattern always showed interference dislocations as a signature of a vortex and its retro-reflected copy image for creating a single vortex. In order to measure clear interference pattern image representing only real vortices, we used and even improved MZ interferometer.

The constant phase is used as an reference image for MZ interferogram in overlapping the images of each arm. For this reference image beam, constant phase region of reference image was magnified. In the process, the intensity was decreased, degrading clarity of interference, and constant phase region also can be distorted in magnification, inducing curvedness in interference. To improve these points, one of the arms was designed to have a uniform phase in Gaussian beam. A local region of beam in that arm goes through a pinhole by spatial filtering and then expanded through lens array. The beam of the other arm had emission of polariton condensate including vortices, and was overlapped with the uniform phase reference beam. This method can remove copied vortex image altogether and also reduce the distortion of interference pattern from local wave front distortion or phase fluctuation. It shows more clearly interference pattern of real vortex much more clearly.

In the experiment, we checked that the repeatability of pulsed pump laser is high enough across pulses to measure time-integrated image averaging over the polariton dynamics. The pulsed pump has a repetition rate of 80 MHz which has the temporal period between each pulse is about 12.5 ns (pulse temporal width = 150 fs). Because our CCD exposure time is more than 100 ms, the ensemble averaged PL signal in pumping with at least 8 million pulse trains was measured on CCD of monochromator and interferometer. The fact that position and phase of polariton vortex wasn't smeared out in time-integrated measurement guarantees a sufficiently high level of reproducibility and repeatability in generation and manipulation of vortex.

We performed the same experiment of the manuscript over many sample locations and the results are all very consistent. We located a vortex by non-resonant pulsed LG pumping and changed the sample position in situ using a piezostage. One can see that the position of the vortex remains almost the same in pumping region relative to the different sample positions [3]. If the vortex was created and pinned on a defect, the vortex should have been dragged along with the sample.

## S II. DISPERSION CURVE WITH ANGULAR MOMENTUM

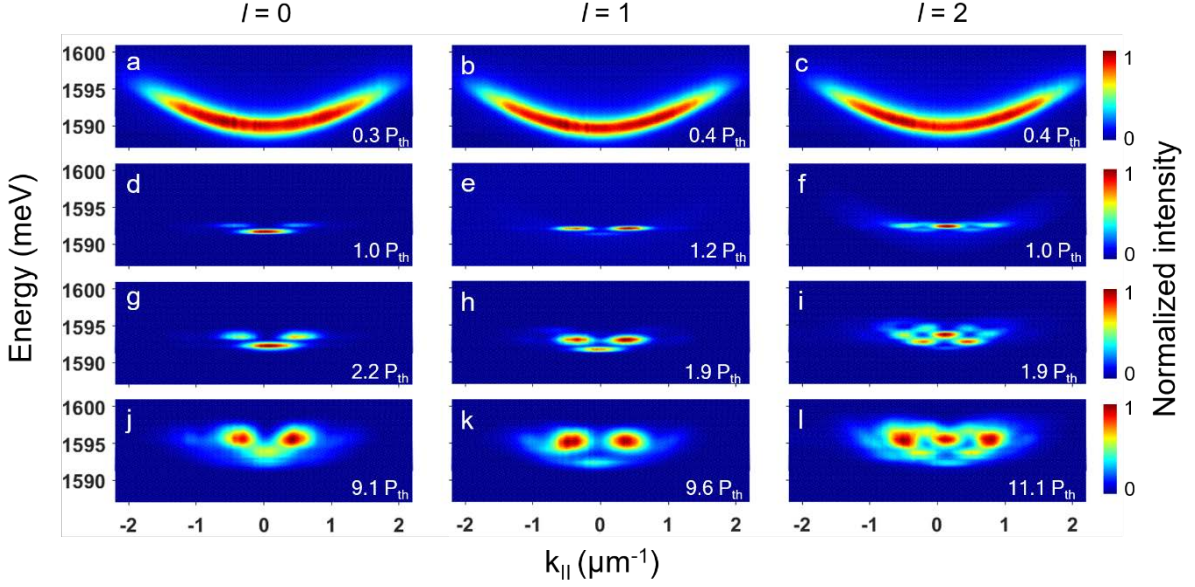

FIG. S2. Power dependence of angle-resolved photoluminescence. (a,d,g,j) Energy-momentum dispersions of the ring polariton condensate under non-resonant pumping without phase winding ($\ell = 0$) at varying pump powers: 0.3 $P_{th}$ in a, 1.0 $P_{th}$ in d, 2.2 $P_{th}$ in g, 9.1 $P_{th}$ in j. (b,e,h,k) Energy-momentum dispersions of $\ell = +1$ vortex under non-resonant pumping with phase winding ($\ell = +1$) at varying pump powers: 0.4 $P_{th}$ in b, 1.2 $P_{th}$ in e, 1.9 $P_{th}$ in h, 9.6 $P_{th}$ in k. (c,f,i,l) Energy-momentum dispersions of two $\ell = +1$ vortices under non-resonant pumping with phase winding ($\ell = +2$) at varying pump powers: 0.4 $P_{th}$ in c, 1.0 $P_{th}$ in f, 1.9 $P_{th}$ in i, 11.1 $P_{th}$ in l. Every image was detected at the same sample region. Due to the numerical aperture (NA = 0.55) of the objective lens, the in-plane momentum range is from -2.2 $\mu m^{-1}$ to 2.2 $\mu m^{-1}$.

To check the energy of the vortex states, we measured the dispersion curve using angle-resolved photoluminescence. Below the threshold power for condensation, regardless of the phasemask presence, all dispersion curves are identically parabolic showing a free-particle-like polariton dispersion, from which the polariton effective mass was extracted and determined to be $8 \times 10^{-5}$ $m_e$. Here $m_e$ is the bare electron mass.

The ground state energies were also found to be identically 1590 meV. (Fig. S2a-c.) Above the threshold power, however, the presence of phasemasks dramatically alters the dispersion curve. Without the vortex formation, one would expect polaritons to condense into the zero momentum state. With vortex formation, the predominant condensed state is a finite momentum state, not the zero momentum state. This can be seen by comparing Fig. S2d-f slightly above the threshold power. In the case of 2 $P_{th}$, the discrete state can be seen more clearly (Fig. S2g-i). Specifically, in the case of $\ell = +1$ and $+2$, it can be seen that a higher state is generated compared to the case of $P_{th}$ (Fig. S2h,i). Also, this discrete state is shown to maintain well when power up to 11 $P_{th}$ is applied (Fig. S2j-l).

In the $\ell = +1$ and $\ell = +2$ cases, the excited state population was larger than the ground state population, in contrast to the $\ell = 0$ case, where the ground state was mostly occupied. For $\ell = +2$, a second excited state appears dominantly right above the threshold power, indicating that the two vortex state carries more energy than a single vortex state, as it should. There is a flowless region in the two vortices states, right in between the two vortices, as a result of counterflow canceling each other out. The zero momentum state with excited energy can be seen in this case.

Meanwhile, polariton confinement due to the ring geometry of the non-resonant pumping ($\ell = 0,+1,+2$) can generate a noticeable feature, in that the excited energy states are discrete [4]. The effective mass of the polariton being $8\times10^{-5}$ $m_e$, a few μm confinement should produce the observed energy gap of around 1.3 meV between the two successive states. This is consistent with our ring-shaped beams having roughly 10 μm diameter.

# SIII. STABILITY OF VORTEX WHILE CHANGING TOWARD THE ASYMMETRIC SPATIAL INTENSITY DISTRIBUTIONS OF THE LG PUMP BEAM

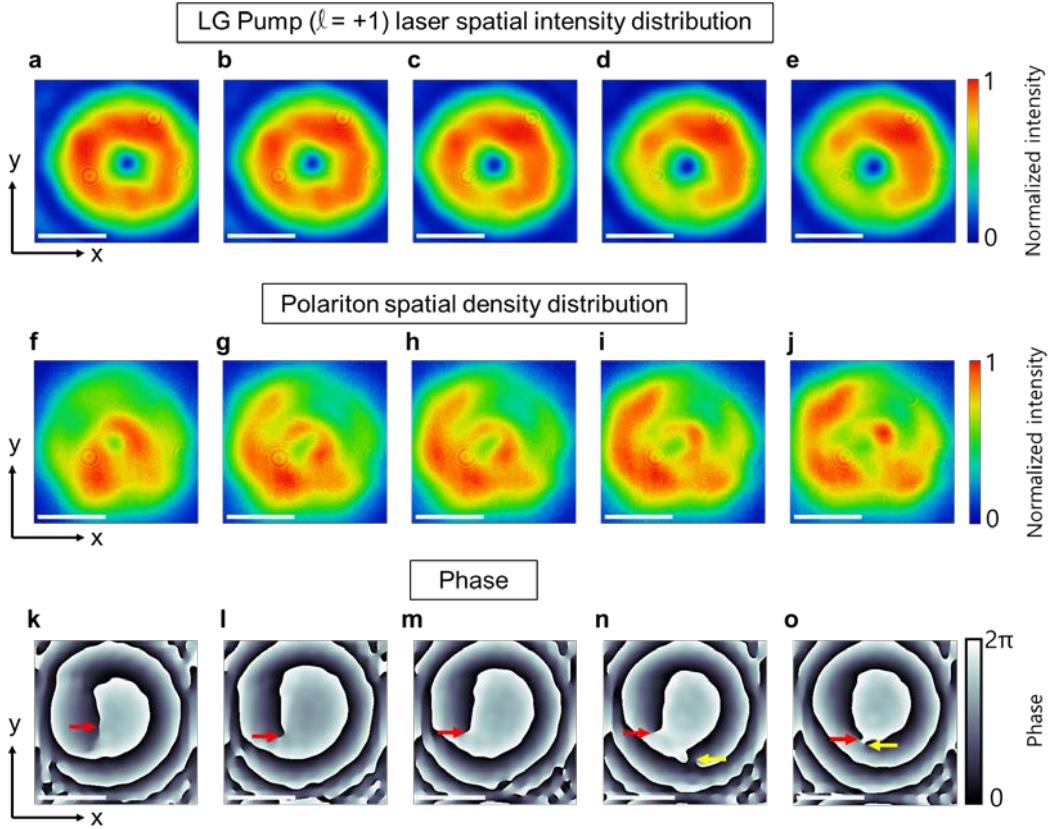

FIG. S3. Stability of the vortex while changing toward the asymmetric spatial distributions of the LG pump (OAM, $\ell = +1$) spatial intensity and polariton density, considering the spatial interference of the polaritons in MZ interferometer. (a-e) The LG pump laser intensity distribution was changed toward a more asymmetric shape from a (ring shape) to e (an asymmetrically collapsed ring shape) by manipulating the spatial position of the phase mask on the pump beam path in the experiment setup. The pump laser beam diameter, about 30 μm. (f-j) The polariton spatial density distribution varied from f (ring shape pumping) to j (asymmetrically collapsed ring shape pumping) toward the asymmetric shape. (k-o) Extracted phase map image of the polariton condensate, corresponding to (f-j). A single vortex (red arrow) was created. (o) Additional anti-vortex moving to inside from outside of pump region in the phase map of the polariton (j). Correspondence relationship: (a→f→k), (b→g→l), (c→h→m), (d→i→n), (e→j→o). These phase images were measured by MZ interferometer. Scale bars, 10 μm.

The stability of the vortex generated by the OAM transfer was further tested using laser beam shapes with a broken circular symmetry. By progressively offsetting the center of the Gaussian beam with respect to the center of the $\ell = +1$ phasemask, the LG pump beam's spatial intensity distribution became asymmetric, as shown in Fig. S3a-e. The pumping power of the Gaussian beam was fixed to $P = 2\, P_{th}$. With a sufficiently large beam spot, the overall OAM of light coming through the phasemask was maintained at $\ell = +1$, up to a point, and only the spatial intensity distribution was affected.

In return, the spatial density distributions of the polaritons varied, as shown from Fig. S3f-j, in response to Fig. S3a-e. In Fig. S3f, when the Gaussian shape pump beam is focused on the center of the phasemask in the experiment setup, polariton density is fairly isotropic, showing a density depleted region in the center.

Fig. S3g-j shows the distortion in the polariton density, causing the ring-shaped distribution to evolve into a horseshoe shape. Fig. S3k-n shows that one counter clockwise phase winding (red arrow in the figure) appears in the vicinity of the polariton density depleted region. When the pump beam shape is distorted to the point of what is shown in Fig. S3e, additional anti vortex moved from outside pumping area was close to the vortex (red arrow). (See Fig. S3o.)

A comparison of the laser intensity (Fig. S3a-e) and the polariton emission (Fig. S3f-j) shows that the polariton emission is brighter in the region where the laser is dimmer. This suggests that polaritons are being pushed away from the high laser intensity region. An asymmetrical pump intensity distribution in real space can weaken the stable trap of the polariton vortex and the transfer of the optical OAM in the center of the LG pumping region.

The normalized intensity distribution can be used to visualize the asymmetrical modification of the LG beams between Fig. S3a and Fig. S3e. When comparing the intensity difference along the ring shape in Fig. S3a (low-intensity imbalance) and Fig. S3e (high-intensity imbalance), it can be seen that as long as the pump intensity imbalance was below 14% along the ring geometry, the vortex survived. The vortex disappeared for an asymmetry greater than 14% (Fig. S3o). From this, we could test how much of the geometrical modification in the spatial intensity distribution of the laser pump with OAM affects the stability of the vortex. These results could be further developed to obtain dynamic control of the vortices by beam shaping of the non-resonant optical pumping.

# SIV. STABILITY OF TRANSFERRING LIGHT'S ORBITAL ANGULAR MOMENTUM AS A FUNCTION OF THE PUMP BEAM RADIUS.

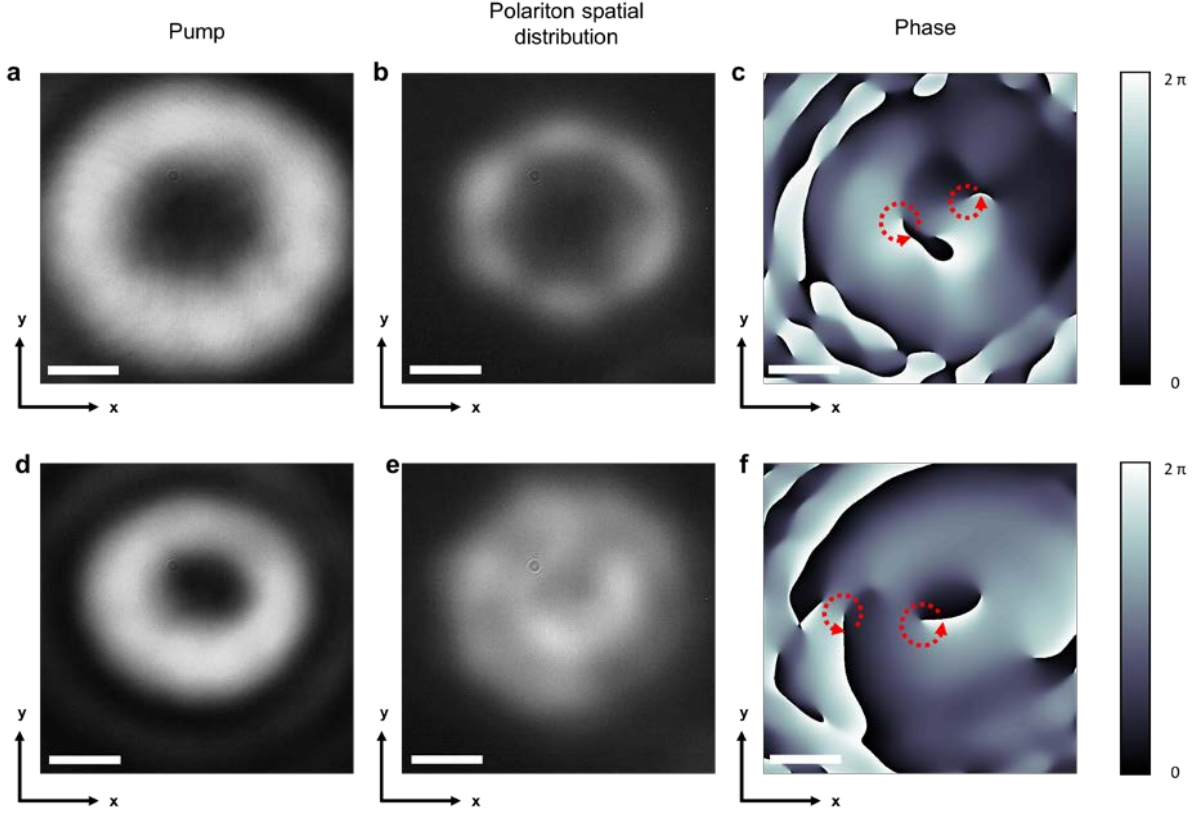

FIG. S4. Stability of transferring the light's orbital angular momentum as a function of the pump beam radii between 7.5 μm (large) and 5 μm (small) with $\ell = +2$ LG beam. Scale bar, 10 μm. (a) The $\ell = +2$ LG pump laser intensity spatial distribution with a 7.5 μm radius. Pump power, 3 mW. (b) Polariton spatial distribution under laser pumping in (a). (c) Polariton phase map image extracted from Michelson interferometer under laser pumping in (a). Red circles indicate the location of real vortices (c, f). Two vortices were generated inside the LG beam pumping region. (d) $\ell = +2$ LG pump laser intensity spatial distribution with 5 μm radius. Pump power, 3 mW. (e) Polariton spatial distribution under laser pumping in (d). (f) Polariton phase map image extracted from Michelson interferometer under laser pumping in (d). One of the two generated vortices was located in the boundary area of the LG beam pumping. The other vortex was located inside the LG beam pumping region. Scale bars, 10 μm.

In order to resolve whether the origin of the vortex formation was the polariton mode selection caused by the spatial pump beam size dependence or the transfer of the optical OAM, the LG pump beam radius was varied from 5 to 7.5 μm. We performed an experiment by changing the pump beam size with a fixed pump power of 3 mW. With an $\ell$ = +2 phase mask, a condensate with two vortices was formed. Fig. S4a is a spatial distribution image of a laser beam with a beam radius of 7.5 μm. And Fig. S4b is a fluorescence (polariton spatial distribution) image and Fig. S4c is the extracted phase map of the polariton spatial interference image. Fig. S4d-f are images of the beam radius of 5 μm, respectively. For a pump beam (doughnut shape) size of roughly 170 μm$^2$, fluorescence from the polariton condensate has an annular shape with a dark region in the middle. Two vortices reside in this central dark region.

However, when we decreased the pump beam size to 95 μm$^2$, the dark region was plugged up by exciton-polaritons. In this case, one of the vortices was pushed out from center to the outside of the region by outward polariton flow, as seen in Fig. S4f. When the beam size was smaller than 95 μm$^2$, only one vortex was seen, since the other one was eventually pushed out of the field of view.

If the vortices had been induced by a flow around the sample defects, the vortices would likely be pinned to the defects. The fact that vortices get pushed around in our experiment supports that there is no defect nearby, and the vortices were induced by the angular momentum of the pump beam.

Also, the vortex number was found to match the laser vortex number in the specific beam size region, even when the beam size was changed. Our result is in stark contrast to a recent theoretical prediction [5]. The model predicted that an annular shaped beam with $\ell$ = 0 can produce quantized vortices by mode selection, in which case the number of vortices should show ring size dependence. However, when we used $\ell$ = +2, we could confirm that the vortex number of the exciton-polariton was robustly maintained in beam sizes ranging from 95 to 170 μm$^2$. This strongly supports that the vortex formation in our experiment was not due to the geometric effect of the incident beam, but rather the OAM of the laser.

## S V. ANGULAR MOMENTUM TRANSFER IN RESONANT PUMP

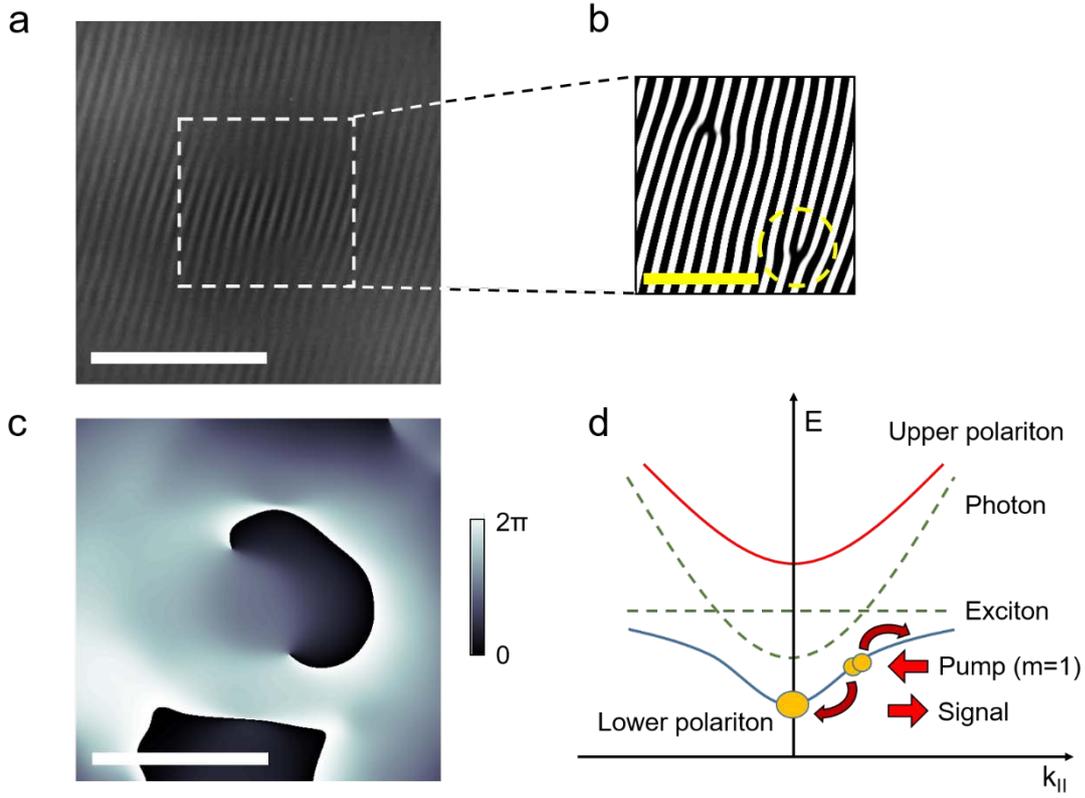

FIG. S5. Creation of a single vortex in an optical parametric oscillator (OPO) pump with $2\pi$ phase winding. (a) Spatial interference of the $\ell = +1$ vortex generated from an OPO pump in the center area of the image in the Michelson interferometer. (b) Magnified image of the dashed square region in a. In the paired fork interference patterns, the one marked by a white circle indicates the real $\ell = +1$ vortex, and the other is an artifact due to the retro-reflector in the Michelson interferometer. (c) Phase map extracted from a, including $2\pi$ phase winding of vortex. (d) OPO pumping conceptual scheme in the energy-momentum dispersion. White scale bar, 10 μm. Yellow scale bar, 5 μm. Pump beam size is 30 μm.

To compare different vortex generating methods, we performed an optical parametric oscillator (OPO) scheme with an $\ell = +1$ phasemask. We injected the pump beam at the inflection point of the lower polariton branch dispersion curve, and the pump beam was a Laguerre-Gaussian beam with the OAM, $\ell = +1$. In this scenario, the polariton pairs scatter to the signal and idler states of the lower polariton branch through polariton-polariton parametric scattering. The pump spot is carefully chosen so that the signal state coincides with the zero momentum ground state of the dispersion curve. Because condensation which is produced by the OPO resonant pump does not experience scattering with the incoherent reservoir, the coherence is known to survive better.

Also, due to the small energy difference between the ground state and the pump state, the resonant OPO pump scheme has little relaxation process from the pump to the final state. The pump beam's OAM is preserved during the resonant scattering process within the lower polariton branch. When the $\ell = +1$ state is scattered into the signal state, we can observe the quantized vortex, as shown in Fig. S5a,b.

Our method of generating vortices with non-resonant pumping has distinct advantages over the OPO technique. The nonresonant method does not require pumping the system with a particular momentum and energy known as the magic angle. It is also easier to separate polariton luminescence from laser reflection, as they have noticeably different wavelengths.

## SVI. Spatial resolved PL

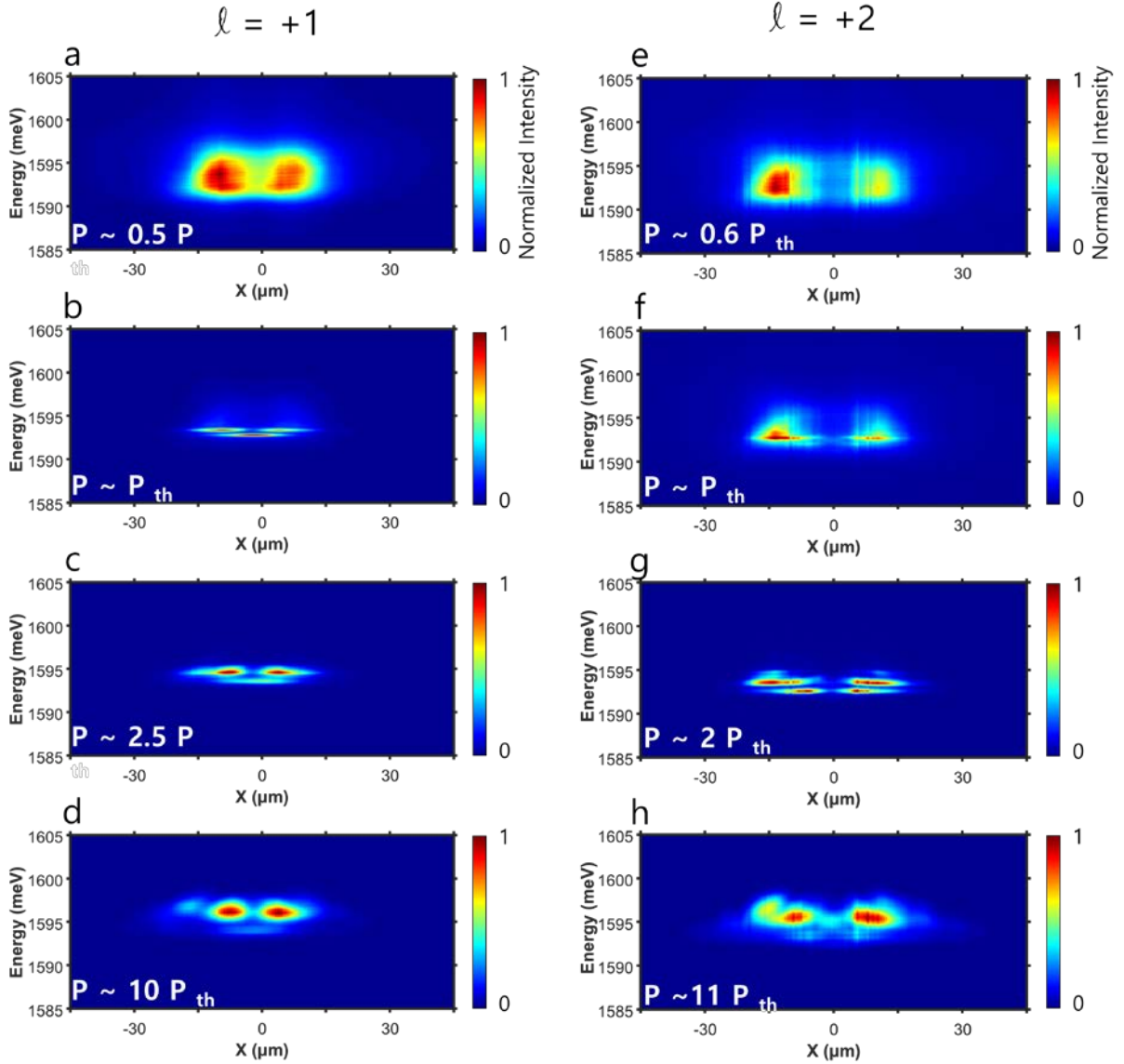

FIG. S6. Measured real space resolved PL of polariton in the each pumping power. (a-d) real space spectra in generation of a single vortex in non-resonant pump with total OAM $\ell = +1$. (e-h) in generation of two vortices in non-resonant pump with total OAM $\ell = +2$

In order to understand the role of optical induced potential and density profile of polaritons, we performed energy resolved photo-luminescence in real space. Under non-resonant LG beam pumping, the reservoir excitons generated from pumping area diffuse around pumped area but cannot approach to the center of the ring shape region due to large effective mass of exciton. Below the threshold density, uncondensed polaritons begin to flow radially inward and outward from the ring shape exciton reservoir due to repulsion with excitons [Fig.S6. a,e]. As pumping power is increased above the threshold density, more polaritons can flow into the center of pump region. The polaritons can build up with large overlap of polariton wavefunctions (long range coherence) and transit to polariton condensates with discrete energy states through stimulated scattering [Fig.S6.b.f][6]. Because polariton condensates experience the confinement due to ring shape exciton reservoir along LG beam pumping region, discrete energy states are formed in the above threshold polariton density. Meanwhile the other fraction of polaritons can create vortex state excited by transferred of optical OAM ($\ell = +1, +2$). In the above threshold density, the quantized rotational motion of collective polaritons contribute to excited energy state inside LG pump region in energy resolved photoluminescence of real space corresponding. As pumping power increase, the discrete energy states undergo blueshift induced by increased interaction between polaritons and exciton reservoir [6]. Higher energy states are relatively enhanced than ground state with zero orbital angular momentum by existence of vortex, as pump power increase [4,7,8].

**SVII. Generation of vortex under non-resonant continuous wave (CW) pumping**

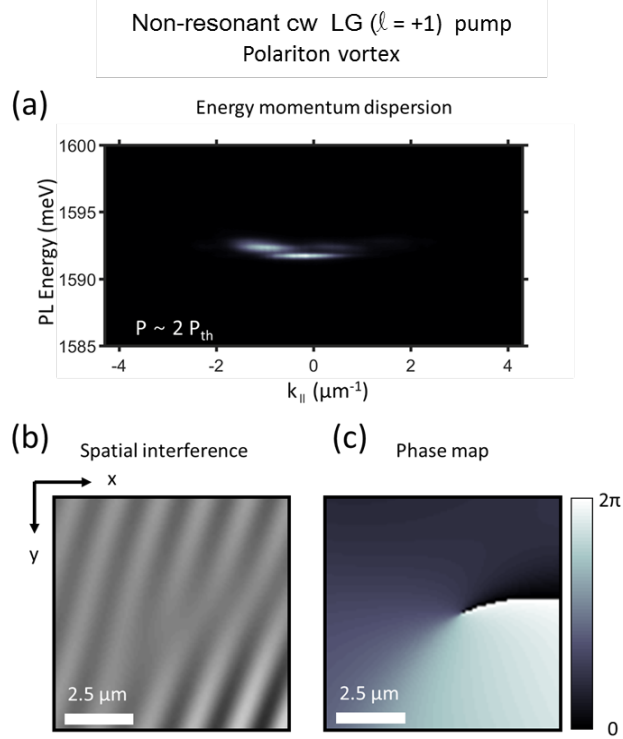

FIG. S7. Generation of single quantized vortex in CW non-resonant LG ($\ell$ = +1) pumping by direct transfer of optical OAM above threshold pumping power at 2 $P_{th}$. (a) Energy momentum dispersion, (b) spatial interference fork pattern and (c) phase map with winding of quantized vortex.

We considered steady-state of polaritons in continuous wave (cw) pumping with the same cryostat temperature and pump beam path. In the experiment setup, non-resonant cw LG ($\ell$ = +1) pump has the same excitation wavelength ($\lambda \sim$ 716 nm) and similar beam diameter size ($\sim$ 30 μm) with pulsed cw LG beam. The cw pump beam was chopped per few microseconds repeatedly for avoiding sample heating, using laser beam chopper. In off-resonant, cw LG ($\ell$ = +1) pump (716 nm), direct transfer of optical OAM could generate a quantized vortex in steady state polariton condensate. Quantized phase winding and spatial interference fork pattern of polariton wavefunction appeared above threshold density [FIG. S7. b,c]. Discrete energy states also arised, indicating collective polariton state, in energy-momentum dispersion [FIG. S7. a]. The cw LG pumping beam could localize and maintain vortex of steady state polariton stationary along pumping area as pulse LG pump beam did. These results can be helpful to explore the steady state polariton quantum fluid.

## SVIII. Polariton vortices as varying the pumping power upto very high intensity.

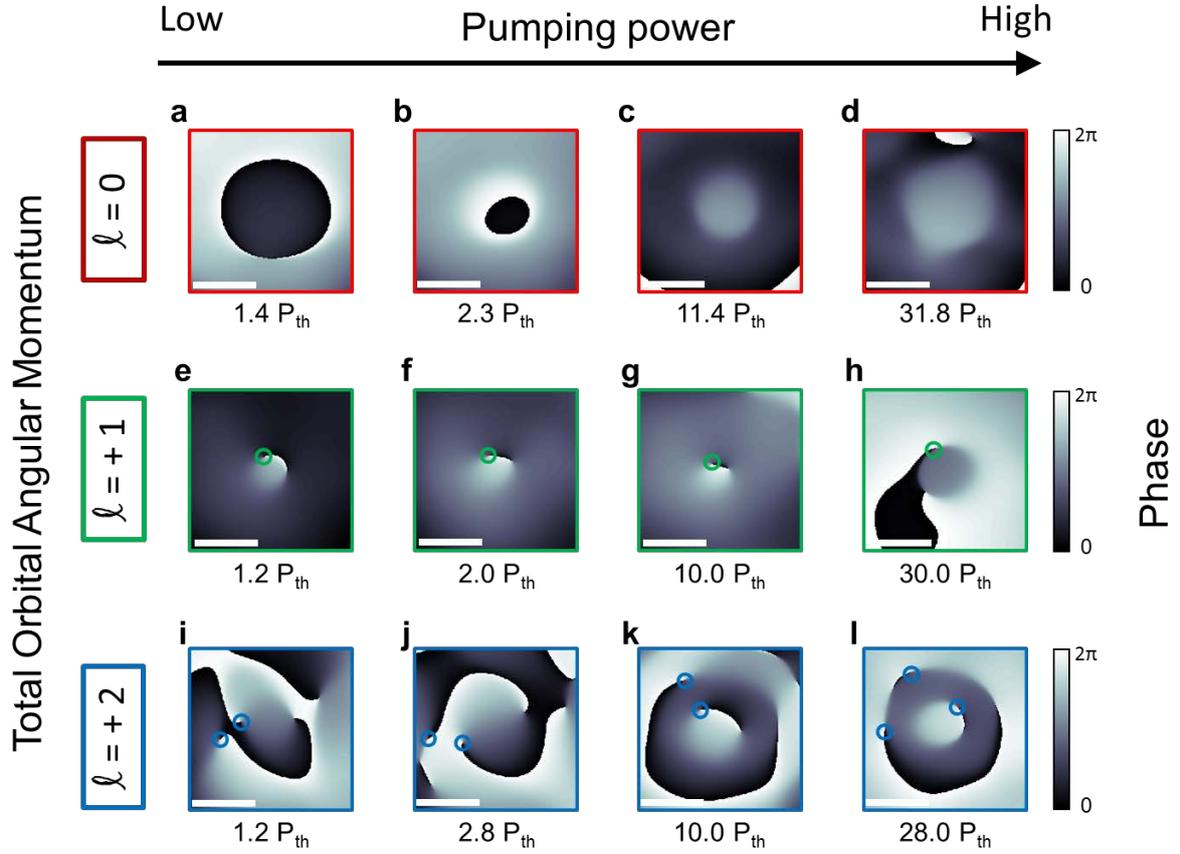

FIG. S8. Phase map of polariton vortices as varying the pumping power. Phase map images of the polariton condensates generated by non-resonant pumping with (a-d) $\ell = 0$, (e-h) $\ell = +1$ and (i-l) $\ell = +2$ are shown for various pumping powers. Phase map images were extracted from Michelson interferogram. (a-d) Polariton condensate ($\ell = 0$) with increasing pump power up to 31.8 $P_{th}$ shows no clear vortex. (e-h) A single vortex ($\ell = +1$) is stable up to pump power of 30.0 $P_{th}$. (i-l) Doubly quantized vortices are stable up to 10.0 $P_{th}$ with $\ell = +2$ LG pump beam. Scale bars, 5 μm. Green circle (Blue) indicates position of single vortex (doubly quantized vortices).